\documentclass[12pt]{article}

\usepackage{geometry}
\usepackage[square,sort,comma,numbers]{natbib}
\usepackage{graphicx}
\usepackage{amsmath,amsfonts,amssymb}
\usepackage[dvipsnames]{xcolor}
\usepackage{braket}
\usepackage{physics}
\usepackage{subfig}
\usepackage{float}

\usepackage{authblk} 

\DeclareMathOperator{\rea}{\textrm{Re}} 

\begin{document}
\title{Temporal evolution of a driven optomechanical system in the strong coupling regime}

\author[1]{L. Medina-Dozal}
\author[1]{J. R\'ecamier}
\author[2]{H. M. Moya-Cessa}
\author[2]{F. Soto-Eguibar}
\author[3]{R. Rom\'an-Ancheyta\thanks{ancheyta@fata.unam.mx}}
\author[2]{I. Ramos-Prieto}
\author[1]{A. R. Urz\'ua}

\affil[1]{ {\small Instituto de Ciencias F\'isicas, Universidad Nacional Aut\'onoma de M\'exico. Avenida Universidad s/n, Col. Chamilpa, Cuernavaca, Morelos, 62210 Mexico.}}

\affil[2]{{\small Instituto Nacional de Astrof\'isica, \'Optica y Electr\'onica, Luis Enrique Erro 1, Santa Mar\'ia Tonantzintla, Puebla, 72840 Mexico}}

\affil[3]{{\small Centro de F\'isica Aplicada y Tecnolog\'ia Avanzada, Universidad Nacional Aut\'onoma de M\'exico, Boulevard Juriquilla 3001, Quer\'etaro 76230, Mexico}}

\date{\today}

\maketitle
\begin{abstract}
We obtain a time-evolution operator for a forced optomechanical quantum system using Lie algebraic methods when the normalized coupling between the electromagnetic field and a mechanical oscillator, $G/\omega_m$, is not negligible compared to one\textcolor{black}{, i.e., the system operates in the strong-coupling regime.} Due to the forcing term, the interaction picture Hamiltonian contains the number operator in the exponents, and in order to deal with it, we approximate these exponentials by their average values taken between initial coherent states. Our approximation is justified when we compare our results with the numerical solution of the number of photons, phonons, Mandel parameter, and the Wigner function, showing an excellent agreement. \textcolor{black}{In contrast to other works, our approach does not use the standard linearized description in the optomechanical interaction. Therefore, highly non-classical (non-Gaussian) states of light emerge during the time evolution.}
\end{abstract}

\section{Introduction}
The simplest system describing the main aspects of cavity optomechanics consists of an optically driven Fabry-P\'erot resonator with one end mirror fixed and the other harmonically bound and allowed to oscillate due to the radiation pressure from the intracavity field. In a standard optomechanical system, the mirror's position parametrically modulates the frequency of the optical cavity mode. Compared with the mechanical frequency of the mirror, $\omega_m$, and the cavity line width, the optomechanical coupling, $G_0$, is usually small~\cite{RMP_2014_Florian, Verhagen2012}. However, the effective coupling is a function of the cavity photons; for many photons, this coupling increases by $\sqrt{n}$ \cite{Meystre2012}. 

\textcolor{black}{
The extension of an optomechanical system to tripartite cavity optomechanical systems by the inclusion of a two-level atom in the optical cavity was considered in \cite{Laha2019}; there, \textcolor{black}{the authors found collapses in the von Neumann entropy of the atom's subsystem} the authors found collapses in the atom's subsystem von Neumann entropy when an intensity-dependent coupling between the field and the two-level atom is used. This coupling is written as a product between the usual creation-annihilation operators and the real function of the photon number operator, similar to the {\it f}-deformed operators introduced by Man'ko and collaborators several years ago \cite{Manko1997}, and whose nonlinear coherent states were analyzed in \cite{JPA_Ancheyta_2011,Santos2011}. In \cite{MedinaDozal2020}, the authors studied a similar tripartite system consisting of a pumped optomechanical cavity and a two-level atom coupled with the intracavity field.}
%
\textcolor{black}{In \cite{Medina2022} an approximate study of the temporal evolution for an optomechanical Jaynes-Cummings model (JCM) incorporating a non-perfect cavity, so that photons may escape, and a two-level atom that may decay spontaneously was considered. The authors made an extension of the treatment presented in \cite{Quang1991} for the study of quantum collapses and revivals in an optical cavity.}

Refs.~\cite{Groblacher2009, Aspelmeyer2010} report the experimental normal mode splitting when the coupling strength between the mechanical oscillator and the cavity field is strong enough ($G_0 \geq  \kappa, \gamma_m$), where $\kappa$ is the cavity amplitude decay rate and $\gamma_m$ the mechanical oscillator decay rate~\cite{Marquardt2007}. Under these circumstances, the mechanical oscillator states become dressed with the interacting photons, and these dressed states are entangled states between the field and the mechanical oscillator, in a similar form as a cavity field and a two-level atom are entangled in the case of the JCM~\cite{Haroche2006}. 

\textcolor{black}{
The temporal evolution of an {\it f}-deformed optomechanical system has been considered recently in \cite{Dehghani2022}. There, the authors introduce nonlinearities in the electromagnetic field using a deformation function and study its relevance in the evolution of the entanglement between the field and the mechanical oscillator calculating the linear entropy and the Wigner function. In \cite{Eftekhari2022}, also considering {\it f}-deformed interactions, the dynamics of a three-level atom with a non-linear atom-field coupling in an optomechanical cavity with a two-mode field is studied and the evolution of several non-classical properties is presented. These kinds of non-linearities have also been applied to the JCM~\cite{Altowyan2020, Rodriguez2013}.}

In Refs.~\cite{Rai2008,Machado2019}, an optomechanical system with a coupling proportional to the squared displacement operator was considered. They obtained that the expected mechanical displacement shows collapses and revivals similar to those in the JCM when the cavity and mechanical oscillator are both in coherent states. Here, we show that without the quadratic nonlinear interaction, a driven optomechanical system can also display the phenomena of collapses and revivals when working in the strong coupling regime. 

In this work, using Hamiltonian parameters consistent with experiments reported in the literature \cite{Verhagen2012, FriskKockum2019,Brennecke_2008,Amir_2014}, we show in detail the temporal evolution of several physical observables \textcolor{black}{of a driven optomechanical system}, like the average value of the photon number, the average value of the phonon number, the linear entropy for the mirror, and the Wigner function for the field and the mechanical oscillator. Our results, obtained with the time evolution operator in the strong coupling regime, are in excellent agreement with those obtained by purely numerical means.

We structure our paper as follows: In Sec.~\ref{sect-theory}, we present a Lie algebraic approach to obtain the exact solution for the undriven optomechanical system. We evaluate the evolution of the average phonon number and find that for strong enough coupling, the mechanical oscillator changes its behavior from cooling to heating, with all the other Hamiltonian parameters fixed. Then, we construct an approximate time-evolution operator for a driven optomechanical system applicable even when the coupling $G_0/\omega_m\sim 1$. In Sec.~\ref{sect-observables}, we evaluate the temporal evolution of several observables, like the average number of photons and phonons, the Mandel parameter, and the linear entropy for a wide range of values of the optomechanical coupling. Sec.~\ref{wigner_section} discusses the corresponding Wigner function of the system. Finally, in Sec.~\ref{sect-conclusions}, we present our conclusions.

\section{Theory}\label{sect-theory}
This section first introduces a Lie algebraic approach to obtain the exact solution for an undriven optomechanical system. Our analysis concentrates on the evolution of the average phonon number, revealing an interesting transition: \textcolor{black}{as the coupling strength $G_0/\omega_{m}$ becomes sufficiently large, the mechanical oscillator shifts from a cooling behavior to a heating one}. We will see that this transition persists even when changing Hamiltonian parameters, emphasizing the intricate interplay between the mechanical and optical aspects of the system. We also develop an approximate time-evolution operator tailored for a driven optomechanical system in scenarios where the normalized coupling $G_0/\omega_m$ is close to unity.

\subsection{Optomechanical system}
Let us consider the basic optomechanical interaction, encompassing a mechanical oscillator's frequency $\omega_m$ and an electromagnetic cavity field with frequency $\omega_c$. The quantum Hamiltonian representing this interaction is~\cite{Aspelmeyer2010}
\begin{equation}\label{hamnoforzado}
{\hat{H}_{opt}} =\hbar \omega_c \hat n +\hbar\omega_m \hat N - \hbar G_0\hat n(\hat b+\hat b^{\dagger}),
\end{equation}
where the optomechanical single-photon coupling strength $G_0$ is given by
\begin{equation}\label{G0_optomechanics}
G_0 =\frac{\omega_c}{L}\left(\frac{\hbar}{2m\omega_m}\right)^{1/2}.
\end{equation}
The cavity length is $L$, and $m$ represents the mass of the mechanical oscillator. The operators $\hat{a}$ ($\hat{b}$) and $\hat{a}^\dagger$ ($\hat{b}^\dagger$) are the annihilation and creation operators of the quantized cavity field (mechanical oscillator), respectively. Thus, $\hat{n}=\hat{a}^\dagger\hat{a}$ and $\hat{N} = \hat{b}^\dagger\hat{b}$ are the number operators of the quantized field and mechanical oscillator, respectively. It is easy to recognize that the set of operators appearing in $\hat{H}_{opt}$ is closed under commutation (see Table~\ref{Table1}); therefore, the evolution operator can be expressed exactly as a product of exponentials~\cite{Wei1964}
\begin{equation}\label{U_opt}
\hat{U}_{opt} = e^{\alpha_1 \hat n}e^{\alpha_2\hat N} e^{\alpha_3 \hat n\hat b^{\dagger}} e^{\alpha_4 \hat n\hat b} e^{\alpha_5 \hat n^2}.
\end{equation}
Upon substituting $\hat{U}_{opt}$ into the Schrödinger equation, we obtain the complex time-dependent functions $\alpha_i$,
\begin{subequations}
\begin{align}
\alpha_1  = & -i\omega_c t,\\
\alpha_2  = &-i \omega_m t,\\
\alpha_3  = &-\frac{G_0}{\omega_m}\left(1-e^{i\omega_m t}\right), \label{009c00} \\
\alpha_4  = &-\alpha_3^*, \\
\alpha_5 = &\left(\frac{G_0}{\omega_m}\right)^{2}(i\omega_m t-1 + e^{-i\omega_m t}).
\end{align}
\end{subequations}
To express the evolution operator, Eq.~\eqref{U_opt}, in the context of the standard displacement operator of quantum optics~\cite{Gerry2004}, we first clarify that 
\begin{equation}
\hat{D}_{\hat x}(\alpha) \equiv e^{\alpha \hat{x}^{\dagger}-\alpha^{*}\hat{x}}= e^{-\frac{1}{2}|\alpha|^2} e^{\alpha\hat{x}^{\dagger}}e^{-\alpha^{*}\hat{x}}
\end{equation}
serves as the representation of the Glauber displacement operator associated with the quantized field and mechanical oscillator (with $\hat x = \hat a, \hat b$), and $\alpha\in\mathbb{C}$. By utilizing the commutation relations provided in Table \ref{Table1}, we can express the evolution operator as
\begin{equation}\label{eq:uopto}
\hat{U}_{opt} =e^{\alpha_2 \hat{N}} e^{\alpha_1 \hat{n}} e^{\alpha_5 \hat{n}^2} e^{\frac{1}{2}|\alpha_3 \hat{n}|^2}\hat{D}_{\hat b}\left(\alpha_3 \hat{n}\right).
\end{equation}
\begin{table}[H]
\centering
\begin{tabular}{||c c c c c c||} 
 \hline
& $\hat{n}$ & $\hat{N}$ & $\hat{n}\hat{b}$ & $\hat{n}\hat{b^\dagger}$ & $\hat{n}^2$ \\ 
\hline\hline
$\hat{n}$ & 0   & 0  & 0                 & 0                        & 0           \\
$\hat{N}$ & 0   & 0   & $-\hat{n}\hat{b}$ & $\hat{n}\hat{b}^\dagger$ & 0           \\
$\hat{n}\hat{b}$& 0  & $\hat{n}\hat{b}$   & 0                 & $\hat{n}^2$          & 0  \\
$\hat{n}\hat{b}^\dagger$ & 0  & $-\hat{n}\hat{b}^\dagger$ & $-\hat{n}^2$   & 0   & 0           \\
$\hat{n}^2$  & 0    & 0       & 0          & 0                        & 0
\\
\hline
\end{tabular}
\caption{Commutation relations associated with the operators within the Hamiltonian, Eq.~\eqref{hamnoforzado}. We have included the operator $\hat n^2$ to account for the commutator between $\hat n\hat b$ and $\hat n\hat b^{\dagger}$.}
\label{Table1}
\end{table}
Now, we choose the initial state to be a tensor product of Glauber coherent states~\cite{Glauber_1963}, encompassing both the field and the mechanical oscillator,
\begin{equation}
\ket{\Psi(0)} = \ket{\alpha}_f \otimes \ket{\Gamma}_m\equiv\ket{\alpha,\Gamma},
\end{equation}
where $\alpha$ and $\Gamma$ are the amplitude of their corresponding coherent state. Using  $\hat{U}_{opt}$, the time-evolution of the coherent states in terms of their corresponding Fock states has the explicit form 
\begin{equation}
\ket{\Psi(t)}= e^{-\frac{1}{2}|\alpha|^2}\sum_{k=0}^\infty \frac{\alpha^k}{\sqrt{k!}}e^{-i[\omega_c t-\Im(\alpha_3\Gamma^{*})]k} e^{ i\left(\frac{G_0}{\omega_m}\right)^2 [\omega_m t-\sin(\omega_m t)] k^2}\ket{k,\Gamma_k(t)},
\end{equation}
with
\begin{equation}\label{Gammak}
\Gamma_k(t) = (\Gamma+k \alpha_3)e^{-i\omega_m t} = \Gamma e^{-i\omega_m t} -k \frac{G_0}{\omega_m}\left(e^{-i\omega_m t}-1\right),
\end{equation}
where the entanglement between the field and the mechanical oscillator is evident. Using this result, we obtain the average number of photons 
\textcolor{black}{
\begin{equation}
\langle \Psi(t)|\hat n|\Psi(t)\rangle = e^{-|\alpha|^2}\sum_{k=0}^\infty\frac{|\alpha|^{2k}}{k!}k = e^{-|\alpha|^2}\sum_{k=1}^\infty \frac{|\alpha|^{2k}}{(k-1)!} =e^{-|\alpha|^2}|\alpha|^2 \sum_{k'=0}^\infty\frac{|\alpha|^{2k'}}{k'!} = |\alpha|^2.
\end{equation}
}
and the average number of phonons
\begin{equation}
\bra{\Psi(t)}\hat{N}\ket{\Psi(t)}=e^{-|\alpha|^2} \sum_{k=0}^\infty \frac{|\alpha|^{2k}}{k!} |\Gamma_k(t)|^2=e^{-|\alpha|^2}\sum_{k=0}^\infty \frac{|\alpha|^{2k}}{k!}|\Gamma+k \alpha_3|^2.
\end{equation}
The summation above for the average number of phonons can be done, and we arrive to
\begin{equation}\label{eq:Naver}
\langle\hat{N}(t)\rangle = |\Gamma|^2 +\left(\alpha_3 \Gamma^{*}+\alpha_3^{*}\Gamma+|\alpha_3|^2\right)|\alpha|^2
+|\alpha_3|^2 |\alpha|^4.
\end{equation}
\color{black}
Using the explicit form for the function $\alpha_3$, Eq. \eqref{009c00}, we obtain for $\Gamma=\Gamma_x+i\Gamma_y$,
\begin{align} \label{eq:phononaver}
\langle \hat N(t)\rangle  = & |\Gamma|^2 
+\left(\frac{2G_0}{\omega_m}\right)^2 \sin^2\left(\frac{\omega_m t}{2}\right)|\alpha|^4
\nonumber \\ &
+\left\{\left[\left(\frac{2G_0}{\omega_m}\right)^2 
-\frac{4G_0}{\omega_m} \Gamma_x \right] \sin^2\left(\frac{\omega_m t}{2}\right)
+\frac{2G_0}{\omega_m} \Gamma_y \sin\left(\omega_m t\right)\right\} |\alpha|^2.
\end{align} 
For $\Gamma$ real ($\Gamma_y=0$), the above expression reduces to
\begin{align}\label{phononavergamareal}
\langle \hat N(t)\rangle  = |\Gamma|^2 +\frac{4G_0|\alpha|^2}{\omega_m^2} \sin^2\Big(\frac{\omega_m t}{2}\Big)\left[G_0(|\alpha|^2+1)-\Gamma\omega_m\right].
\end{align} 

\begin{figure}[H]
\begin{center}
\includegraphics[width=.95\linewidth]{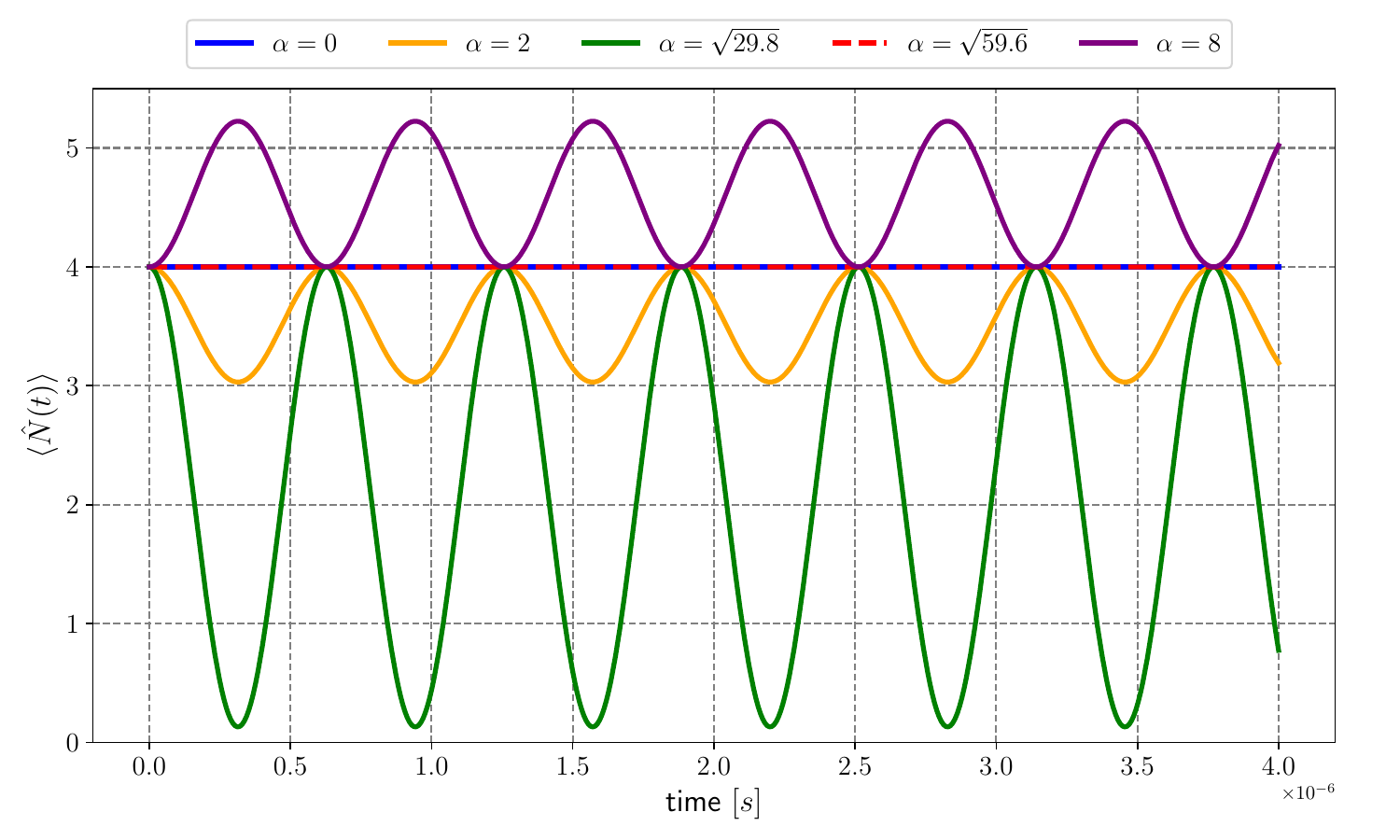}
\caption{Temporal evolution of the average phonon number, \eqref{phononavergamareal}, with parameters $\omega_m= 10^{7}\, \text{s}^{-1}$, $\Gamma=2$, $G_0/\omega_m=0.033$, and the effective coupling is given by $\left(G_{0}/\omega_{m}\right)\vert\alpha\vert^{2}$. The graphics correspond to: $\alpha=0$ (continuous blue), $\alpha=2$ (continuous orange), $\alpha=\sqrt{29.8}$ (continuous green), $\alpha=\sqrt{59.6}$ (dashed red), and $\alpha=8$ (continuous purple).}
\label{figura2}
\end{center}
\end{figure}

We see that the average number of photons in the cavity remains constant, i.e., $\langle \hat{n} \rangle=\left| \alpha\right|^2$, while the average number of phonons of the mechanical oscillator, $\langle\hat{N}(t)\rangle$, depends upon the coupling constant $G_0/\omega_m$, the mean photon number in the cavity $|\alpha|^2$, and the interaction time.
In fact, from the equations \eqref{eq:Naver}, \eqref{eq:phononaver} and \eqref{phononavergamareal}, we can clearly see that the average value of the phonon number of the mechanical oscillator, $\langle\hat{N} (t)\rangle$, depends on the average value of the number of photons in the cavity, $\langle \hat{n} \rangle=\left| \alpha\right|^2$,  quadratically. In the case where $\Gamma$ is real, for values of $\left| \alpha\right|^2$ between $0$ and $\omega_m \Gamma/G_0-1$ the mechanical oscillator cools down, reaching its maximum cooling for the value of $\left | \alpha_\text{min}\right|^2=\frac{1}{2}\left(\omega_m \Gamma/G_0-1\right)$. For values of $\left| \alpha\right|^2>\omega_m \Gamma/G_0-1$ the mechanical oscillator heats up. In this case, that is, with real $\Gamma$, all these values are independent of time. In Fig.~\ref{figura2},  we explicitly show this behavior; we have plotted the mean value of the phonon number of the mechanical oscillator as a function of time for several values of $\alpha$; for $\alpha=0.0$, blue dotted line; for $\alpha=2.0$, orange continuous line; for $\alpha=\sqrt{29.8}$, which is the value of $\alpha$ where the maximum cooling is obtained, green continuous line; for $\alpha=\sqrt{59.6}$, which is the value of $\alpha$ where the original number of phonons is conserved (on average, no new phonons are produced) are obtained, red dashed line; and for $\alpha=8.0$, purple continuous line. For the initial function of the mechanical oscillator we use $\Gamma=2$, and the values of the Hamiltonian parameters are $\omega_m= 10^{7}\, \text{s}^{-1}$, and $G_0 /\omega_m=0.033$.
Analyzing the evolution operator \eqref{eq:uopto}, we can notice that there is an explicit Kerr-type term in it; therefore, from a physical point of view, we can expect the field inside the cavity to evolve in a similar way as it would in a non-linear Kerr medium \cite{Ramos_2021,yurke86,Mancini_1997,Bose1997}. Given the large number of non-classical states that can be produced in a Kerr-type medium, it is evident that this system offers the possibility of producing non-classical states of light. In fact, it is known that like Kerr-type nonlinearity, this system consisting of a cavity with a moving mirror also exhibits optical bistability \cite{Meystre:85}. Therefore, the non-monotony behavior of $\langle \hat N(t)\rangle$ in Fig.~\ref{figura2} is another consequence of the nonlinear response of the optomechanical system.

\textcolor{black}{It is important to note that, in principle, the magnitude of the cavity optomechanical coupling constant $G_0$ given in Eq.~(\ref{G0_optomechanics}) can be increased by changing the parameters, either by increasing the cavity frequency, decreasing the distance between the cavity mirrors or decreasing the mass of the mechanical oscillator. In practice, all these can be done (although challenging) by reducing the size (miniaturizing) of the setup. Values of $G_0$ on the order of MHz have been obtained in micrometer-sized devices~\cite{RMP_2014_Florian}. Besides, there are other experimental platforms where the single-photon strong coupling is pursued, for instance, in the BEC-cavity optomechanical system~\cite{Brennecke_2008}. Toroidal microcavities consistently maintain high-quality cavity and mechanical modes, presenting a significant potential for achieving single-photon strong coupling~\cite{Verhagen2012}. Another promising platform to reach this goal is the optomechanical crystal cavity~\cite{Amir_2014}.}

\color{black}

\subsection{Forced optomechanical system}
Now consider the case where the optomechanical system is driven by a field with amplitude $\Omega$ and with frequency $\omega_p$ of the form
\begin{equation}
\hat V = \hbar \Omega \cos(\omega_p t) (\hat a +\hat a^{\dagger}).
\end{equation}
The total Hamiltonian of the driven optomechanical system is
\begin{equation}\label{Hforced}
    \hat H = \hat H_{opt}+ \hat V,
\end{equation}
and the corresponding time evolution operator is
\begin{equation}
\hat U = \hat U_{opt} \hat U_I,
\end{equation}
with $\hat U_{opt}$ given in Eq.~\eqref{eq:uopto}, and $\hat U_I$ the time-evolution operators in the interaction picture that satisfies
\begin{equation}
i\hbar \frac{\partial \hat U_I}{\partial t} = \left[ \hat U_{opt}^{\dagger} \hat V \hat U_{opt}\right] \hat U_I,
\end{equation}
with the initial condition $\hat U_I(0)=1$.

Transforming the operators $\hat{a}$ and $\hat{a}^\dagger$, one obtains
\begin{equation}
\hat U_{opt}^{\dagger} \hat a \hat U_{opt} = e^{i E(t)(2\hat n+1)}e^{iF(t)(\hat b^{\dagger}e^{i\omega_m t/2}+\hat b e^{-i\omega_m t/2})}\hat a e^{-i\omega_c t},
\end{equation}
and
\begin{equation}
U_{opt}^{\dagger} \hat a^{\dagger} U_{opt} = \hat a^{\dagger}e^{i\omega_c t} e^{-iF(t)(\hat b^{\dagger}e^{i\omega_m t/2}+\hat b e^{-i\omega_m t/2})} e^{-iE(t)(2\hat n+1)},
\end{equation}
where the functions $E(t)$ and $F(t)$ are
\begin{equation}
    E(t) = \left(\frac{G_0}{\omega_m}\right)^2 (\omega_m t -\sin(\omega_m t)),
\end{equation}
and
\begin{equation}
    F(t) = \frac{2G_0}{\omega_m}\sin(\frac{\omega_m}{2}t).
\end{equation}

In reference~\cite{Paredes2020}, the case where $G_0/\omega_m = 0.0236$ was considered, and these functions were simply neglected $F(t)\simeq 0$, $E(t)\simeq 0$; then, it was possible to obtain an approximate time evolution operator whose results were corroborated comparing them with a purely numerical calculation using the corresponding full Hamiltonian. 
In this approximation, valid for $G_0/\omega_m\ll 1$, the interaction Hamiltonian in the interaction picture is
\begin{equation}
\hat H_I = \hbar \Omega \cos(\omega_p t)\left[ \hat a^{\dagger} e^{i\omega_c t}+\hat a e^{-i\omega_c t} \right]\simeq \frac{\hbar\Omega}{2}\left[\hat a^{\dagger} e^{i(\omega_c-\omega_p)t}+\hat a e^{-i(\omega_c-\omega_p)t} \right],
\end{equation}
where we have neglected terms that oscillate rapidly. The corresponding time evolution operator is
\begin{equation}\label{UIRWA}
\hat U_I = e^{\beta_1^{(0)}\hat a^{\dagger}}e^{\beta_2^{(0)}\hat a}e^{\beta_3^{(0)}},
\end{equation}
with $\beta_1^{(0)}(t)=(\Omega/2\Delta)(e^{i\Delta t}-1)$, and the detuning $\Delta=\omega_p-\omega_c$. When $\Delta\rightarrow 0$, $\beta_1^{(0)}=i\Omega t/2$. Note that $\beta_1^{(0)}$oscillates with the frequency of the detuning.

When we do not consider the weak coupling limit, the interaction Hamiltonian is given by
\begin{align}\label{eq:Hinteraction}
\hat H_I  = & \hbar \Omega \cos(\omega_p t)\left[\hat a^{\dagger}e^{i\omega_c t} e^{-iF(t)(\hat b^{\dagger}e^{i\omega_m t/2}+\hat b e^{-i\omega_m t/2})} e^{-iE(t)(2\hat n+1)} \right] 
\nonumber \\ 
& + \hbar \Omega \cos(\omega_p t)\left[e^{i E(t)(2\hat n+1)}e^{iF(t)(\hat b^{\dagger}e^{i\omega_m t/2}+\hat b e^{-i\omega_m t/2})}\hat a e^{-i\omega_c t}\right]. 
\end{align}
It is essential to say that the operators in $\hat H_I$ still form a time-dependent Lie algebra; this was noted first in~\cite{Ancheyta_JOSAB_2017}.
In order to treat the operators in the exponents, at least in an approximate form, we take the average value of the exponentials between initial coherent states for the field and the mechanical oscillator respectively; that is, we take
\begin{align}\label{eq:aproxima}
e^{-iF(t)(\hat b^{\dagger}e^{i\omega_m t/2}+\hat b e^{-i\omega_m t/2})} e^{-iE(t)(2\hat n+1)}  \simeq & ~ _m\langle \Gamma|e^{-iF(t)(\hat b^{\dagger}e^{i\omega_m t/2}+\hat b e^{-i\omega_m t/2})}|\Gamma\rangle_m 
\nonumber \\ & 
\times~  _f\langle \alpha|e^{-iE(t)(2\hat n+1)}|\alpha\rangle_f,
\end{align}
for an initial state $|\Psi(0)\rangle = |\alpha,\Gamma\rangle$. 
The first term in the previous equation is simply
\begin{equation}
_m\langle \Gamma|e^{-iF(t)(\hat b^{\dagger}e^{i\omega_m t/2}+\hat b e^{-i\omega_m t/2})}|\Gamma\rangle_m = e^{-\frac{1}{2}F^2(t)}e^{-iF(t)(\Gamma^{*}e^{i\omega_m t/2}+\Gamma e^{-i\omega_m t/2})},
\end{equation}
and the second one is
\begin{equation}
_f\langle \alpha|e^{-iE(t)(2\hat n+1)}|\alpha\rangle_f = e^{-iE(t)}e^{|\alpha|^2(e^{-2iE(t)}-1)},
\end{equation}
where we have used the fact that $|\alpha\rangle_f$ is not an eigenstate of the photon number operator. 

\begin{figure}[H]
\begin{center}
\includegraphics[width = 0.9\textwidth]{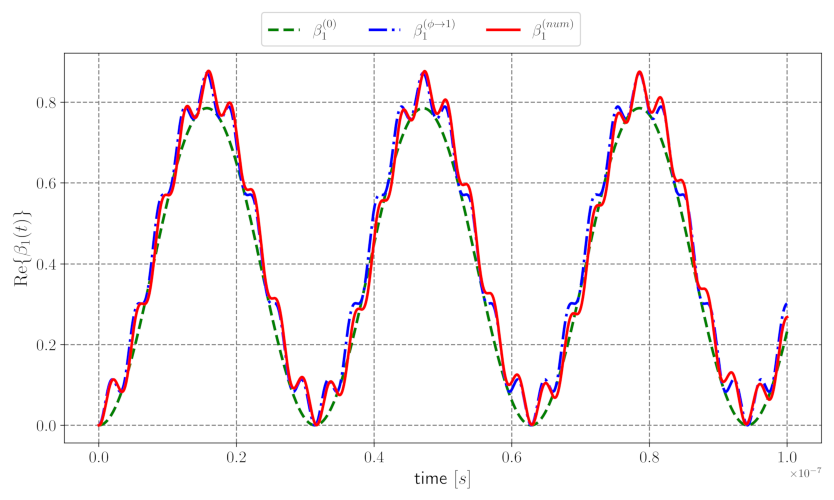}
\caption{Temporal evolution of the real part of function $\beta_{1}(t)$ in the three forms explained above: (Red) $\beta_{1}^{(num)}$, numerical integration of \eqref{betas}; (Blue) $\beta_{1}^{(\phi\rightarrow 1)}$, limiting case when $\phi(t)\rightarrow 1$ in \eqref{betaphi1}; (Green) $\beta_{1}^{(0)}$, in the RWA given by \eqref{UIRWA}. The Hamiltonian parameters used are: $\omega_c=10^9 s^{-1}$, $\omega_m=\omega_c/100$, $\Omega= (\pi/20)\omega_c$, $\omega_p=0.8\omega_c$,  that are required for $\beta_{1}^{(0)}$ and $\beta_{1}^{(\phi\rightarrow 1)}$; whereas, $\beta_{1}^{(num)}$ also requires the values $G_{0}/\omega_{m} = 0.033$, $\Gamma=2$, $\alpha$ = 2.}
\label{beta1}
\end{center}
\end{figure}

In this approximation, the interaction Hamiltonian takes the form
\begin{equation}\label{HApprox}
    \hat H_I = \hbar \Omega \cos{(\omega_p t)}\left[\phi(t) \hat a^{\dagger}e^{i\omega_c t}+\phi^{*}(t) \hat a e^{-i\omega_c t}\right],
\end{equation}
with the time-dependent function
\begin{equation}
\phi(t) = e^{-\frac{1}{2}F^2(t)}e^{-iF(t)(\Gamma^{*}e^{i\omega_m t/2}+\Gamma e^{-i\omega_m t/2})}e^{-iE(t)}e^{|\alpha|^2(e^{-2iE(t)}-1)},
\end{equation}
and the corresponding time-evolution operator given by
\begin{equation}\label{eq:tevol1}
    \hat U_I = e^{\beta_1 \hat a^{\dagger}} e^{\beta_2 \hat a} e^{\beta_3},
\end{equation}
where
\begin{subequations}\label{betas}
\begin{align}
\dot \beta_1  = &  -i \Omega\phi(t) \cos (\omega_{p} t) e^{i\omega_{c} t}, \label{eqs0350}
\\
\dot \beta_2  = & -i \Omega \phi^{*}(t) \cos(\omega_{p}t) e^{-i\omega_{c} t}, \label{eqs0360}
\\
\dot \beta_3  = & -i \Omega \beta_1 \phi^{*}(t) \cos(\omega_{p}t) e^{-i\omega_{c} t}. \label{eqs0370}
\end{align}
\end{subequations}
It can be seen that $\beta_1 = -\beta_2^{*}$, as a consequence of the unitarity of the time evolution operator $\hat U_I$. Due to the product between $\cos(\omega_{p} t)$ and $e^{i \omega_{c} t}$, we can perform a rotating wave approximation (RWA) on this set of equations, whose integration leads to terms like the $\beta_{1}^{(0)}$ introduced above. Also, in the limiting case $\phi(t) \rightarrow 1$ for an arbitrary $t$, we can obtain a behavior that is independent of the coupling; in this sense, the direct integration of $\dot{\beta}_{1}$ gives,
\begin{equation}\label{betaphi1}
\beta_1(t)= \frac{\Omega}{\omega_p^2-\omega_c^2}\left[e^{i\omega_c t}(\omega_c \cos(\omega_p t)-i \omega_p \sin(\omega_p t))-\omega_c\right],
\end{equation} 
where besides the oscillations with the frequency of the detuning, there will be fast oscillations with frequency $\omega_{c} + \omega_{p}$. Fig.~\ref{beta1} shows the evolution of $\Re(\beta_{1}(t))$ in the three forms presented: for the rotating wave approximation in \eqref{UIRWA}, the limit case when $\phi\rightarrow 1$ in \eqref{betaphi1}, and the numerical integration of \eqref{betas}. We see that the numerical integration and the limiting case are nearly overlapped, and present oscillations bounded by the rotating wave approximation curve.
Since $\beta_1=-\beta_2^{*}$, the time evolution operator given by Eq.~\eqref{eq:tevol1} can be written as a displacement operator
\begin{equation}
\hat U_I = e^{\beta_3} e^{\frac{1}{2}|\beta_1|^2} \hat D_{\hat a}(\beta_1),
\end{equation}
and the full-time evolution operator is then
\begin{equation}\label{total_U}
\hat U = e^{-i\omega_c t \hat n} e^{-i\omega_m t \hat N} e^{\alpha_5 \hat n^2} e^{\frac{1}{2}|\alpha_3\hat n|^2} \hat D_{\hat b}(\alpha_3 \hat n) \hat D_{\hat a}(\beta_1)e^{\frac{1}{2}|\beta_1|^2} e^{\beta_3}.
\end{equation}
For an initial state given by the tensor product of a field coherent state and mechanical coherent state, $\ket{\alpha,\Gamma}$, the state vector at time $t$ can be expressed as
\begin{align}\label{wavefunc}
|\Psi(t)\rangle =&  e^{(\beta_3+\frac{1}{2}|\beta_1|^2)}e^{i \Im(\beta_1 \alpha^{*})}e^{-\frac{1}{2}|\beta_1+\alpha|^2} \nonumber \\ & \times
\sum_{k} e^{-i(\omega_c t-\Im(\alpha_3 \Gamma^{*}))k}e^{i\left(\frac{G_0}{\omega_m}\right)^2(\omega_m t-\sin(\omega_m t))k^2}
\frac{(\beta_1+\alpha)^{k}}{\sqrt{k!}}|k,\Gamma_{k}(t)\rangle,
\end{align}
where $\Gamma_{k}(t)$ is defined by the Eq.~\eqref{Gammak}.

\section{Evaluation of observables}\label{sect-observables}
In this section, we evaluate the temporal evolution of several observables, like the average number of photons and phonons, the Mandel parameter, and the linear entropy of the field and the mechanical oscillator.

\subsection{Average number of photons and phonons}
The photon number operator in the Heisenberg representation is
\begin{equation}\label{eq:photonsHeis}
\hat n(t) = \hat U^{\dagger}\hat n \hat U = \hat n -\beta_2 \hat a +\beta_1 \hat a^{\dagger} -\beta_1 \beta_2,
\end{equation}
and the average number of photons is given by
\begin{equation}
\langle \hat{n}(t)\rangle = \langle \alpha| \hat n(t)|\alpha\rangle = |\alpha|^2 -\alpha \beta_2 +\alpha^{*}\beta_1-\beta_1 \beta_2.
\end{equation}
Using that $\beta_1=-\beta_2^{*}$, we obtain
\begin{equation}\label{eq:naver}
\langle \hat n(t)\rangle= |\alpha +\beta_1|^2.
\end{equation}
For the case $G_0/\omega_m\ll 1$, the function $\phi(t)\rightarrow 1$, and we can integrate Eq.~\eqref{betas} to obtain the average value of the photon number as
\begin{equation}\label{eq.0450}
\langle \hat n(t)\rangle = |\alpha|^2+\frac{\Omega}{\Delta}
\left[1-\cos(\Delta t)\right]
\left(\frac{\Omega}{2\Delta} + \rea(\alpha)\right).
\end{equation}
For those cases with larger values of $G_0/\omega_m$, the function $\phi(t)$ is a complicated function of time and we cannot obtain an analytic solution of Eq.~\eqref{betas}.

The average number of phonons is
\begin{equation}\label{eq:phonons}
\langle \hat N(t)\rangle = \langle \Psi(t)|\hat N |\Psi(t)\rangle,
\end{equation}
which has the explicit form
\begin{align}\label{eq:Naverage}
\langle \hat N(t)\rangle= &  e^{2\Re(\beta_3+\alpha\beta_2)}e^{-|\alpha|^2}\sum_{k}\frac{\alpha^k}{k!}
\sum_{p} \frac{\beta_{1}^p}{p!}(k+p)! 
\nonumber \\ & \times
\sum_{k'} \frac{\alpha^{*k'}}{k'!}\sum_{p'} \frac{\beta_1^{*p'}}{p'!}|\Gamma_{k+p}(t)|^2,
\end{align}
with the condition $k+p=k'+p'$. As we show below, the summations can be done analytically.

Starting from Eq.~\eqref{eq:Naverage}, and applying the condition $k+p=k'+p'$,
the third summation can be done, arriving at
\begin{equation}
\langle \hat N(t)\rangle =  e^{2\Re(\beta_3+\alpha\beta_2)}e^{-|\alpha|^2}\sum_{k}\frac{\alpha^k}{k!}
\sum_{p} \frac{\beta_{1}^p}{p!} |\Gamma_{k+p}(t)|^2 (\alpha^*+\beta_1^*)^{k+p}.
\end{equation}
Setting $n=k+p$, then $p=n-k$, we get
\begin{equation}
\langle \hat N(t)\rangle  =   e^{2\Re(\beta_3+\alpha\beta_2)}e^{-|\alpha|^2}\sum_{k}\frac{\alpha^k}{k!}
\sum_{n=k} \frac{\beta_{1}^{(n-k)}}{(n-k)!} |\Gamma_{n}(t)|^2 (\alpha^*+\beta_1^*)^{n}.
\end{equation}
The second sum may be started at zero, as only zeros are added, and the order of the sums can be changed, to get
\begin{equation}
\langle \hat N(t)\rangle =  e^{2\Re(\beta_3+\alpha\beta_2)}e^{-|\alpha|^2}\sum_{n=0}  \frac{|\Gamma_{n}(t)|^2 (\alpha^*+\beta_1^*)^{n}}{n!}\sum_{k}^n\frac{\beta_{1}^{(n-k)}}{(n-k)!}\frac{\alpha^k}{k!}n!,
\end{equation}
that gives
\begin{equation}
\langle \hat N(t)\rangle=e^{2\Re(\beta_3+\alpha\beta_2)}e^{-|\alpha|^2}\sum_{n=0}  \frac{|\Gamma_{n}(t)|^2 |\alpha+\beta_1|^{2n}}{n!}.
\end{equation}
We have $|\Gamma_n|^2=|\Gamma|^2+|\alpha_3|^2n^2+(\Gamma\alpha_3^*+\alpha_3\Gamma^*)n$, so that
\begin{align}
\sum_n \frac{|\alpha+\beta_1|^{2n}}{n!}|\Gamma_n|^2 = & \sum_{n=0}  \frac{ |\alpha+\beta_1|^{2n}}{n!}[|\Gamma|^2+|\alpha_3|^2 n^2+(\Gamma\alpha_3^*+\alpha_3\Gamma^*)n] 
\nonumber \\
=& |\Gamma|^2e^{|\alpha+\beta_1|^{2}} +  (\Gamma\alpha_3^*+\alpha_3\Gamma^*)|\alpha+\beta_1|^2e^{|\alpha+\beta_1|^{2}}
\nonumber \\ & 
+\alpha_3|^2(|\alpha+\beta_1|^2+|\alpha+\beta_1|^4)e^{|\alpha+\beta_1|^{2}},
\end{align}
and using Eq.~\eqref{eq:naver}, we find
\begin{align}\label{eq:schr}
\langle \hat N(t)\rangle = &  e^{2\Re(\beta_3+\alpha\beta_2)}e^{-|\alpha|^2}e^{|\alpha+\beta_1|^{2}} 
[|\Gamma|^2+2\Re(\Gamma\alpha_3^*) \langle \hat n(t)\rangle \nonumber \\ &
+|\alpha_3|^2(\langle \hat n(t)\rangle+\langle\hat n(t)\rangle^2)].
\end{align}
On the other hand, the phonon number operator in the Heisenberg representation is
\begin{equation}
\hat N(t) = \hat N +(\alpha_3 \hat b^{\dagger}+\alpha_3^{*}\hat b)\hat n(t) +|\alpha_3|^2 \hat n(t)^2,
\end{equation}
and taking its average value between coherent states, we obtain
\begin{equation}\label{eq:heis}
\langle \hat N(t)\rangle = |\Gamma|^2 +2\Re(\alpha_3\Gamma^{*})\langle \hat n(t)\rangle +|\alpha_3|^2 (\langle \hat n(t)\rangle +\langle \hat n(t)\rangle^2).
\end{equation}
Comparing Eqs. \eqref{eq:schr} and \eqref{eq:heis}, we see that $2\Re(\beta_3+\alpha\beta_2)-|\alpha|^2 +|\alpha+\beta_1|^2 =0$ as a condition of unitarity; this fact has been confirmed numerically.\\

\begin{figure}[H]
\begin{center}
\includegraphics[width = 0.9\textwidth]{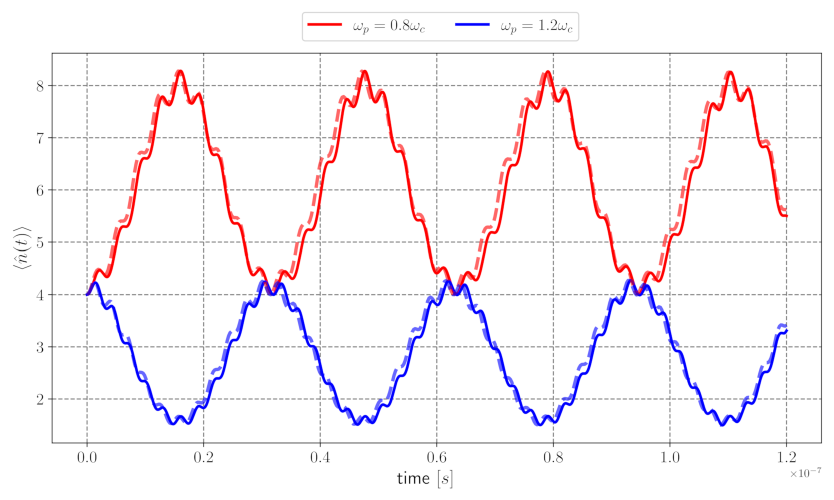}
\caption{Temporal evolution of the average photon number with Hamiltonian parameters  $\omega_c=10^9 s^{-1}$, $\omega_m=\omega_c/100$, $\Gamma=2$, $\alpha$ = 2, $G_0/\omega_m=0.033$, $\Omega= (\pi/20)\omega_c$. Red detuned, $\omega_p=0.8\omega_c$. Blue detuned, $\omega_p=1.2\omega_c$. The analytical expression \eqref{eq.0450} gives the solid lines, whereas the dashed lines are obtained by solving the original Hamiltonian numerically.}
\label{figura3-1a}
\end{center}
\end{figure}

In Fig.~\ref{figura3-1a}, we show the temporal evolution for the average photon number; in solid lines, the results obtained with the method described in the text, and in dotted lines the numerical results obtained using the full Hamiltonian. In these cases, the coupling between the field and the mechanical oscillator is small, $G_0/\omega_m= 0.033$, corresponding to the set of Hamiltonian parameters used in Ref.~\cite{Paredes2020}. The amplitude for the forcing term is $\Omega= (\pi/20)\omega_c$. In red we show the case with $\omega_p=0.8\omega_c$ so that the detuning $\Delta=\omega_p-\omega_c= -0.2\omega_c$ (red detuned) and in blue the case with $\omega_p=1.2\omega_c$ so that $\Delta=0.2\omega_c$ (blue detuned); thus, we are working under non-resonant conditions. Note the almost perfect agreement between the analytic and the numerical calculations for the case of the average photon number.  The initial average photon number is $\bar n= |\alpha|^2 =4$, and for the red detuned case we see that the average photon number is an increasing oscillatory function of time, attaining values in the range $4\leq \bar n(t)< 9$ with a period $T_{\Delta}=(2\pi/\Delta)= 5 T_c$, being $T_c=2\pi/\omega_c$ the period of the cavity. The superimposed fast oscillations with small amplitude have the period $T_{fast}=2\pi/(\omega_c+\omega_p)\simeq T_c/2$ for $\omega_p=0.8\omega_c$. For the blue detuned case, we see that the average photon number is a decreasing oscillatory function of time, its values in the range $1.5 <\bar n < 4.5$; the period of the fast oscillations is again $T_{fast}$, and that of the enveloping is $T_{\Delta}$, as before.  The effective coupling evolves in time according to $\left(G_0/\omega_{m}\right)\langle \hat n(t)\rangle$ and attains values as large as $\simeq 0.25$. \textcolor{black}{It is clear that the increase or decrease in the average photon number depending on the sign of the detuning is a consequence of the behavior of the real part of the function $\beta_1(t)$ which, for the red detuned case takes positive values while for the blue detuned case takes negative values. When the pumped field is red-detuned, the only way to actually scatter photons into the cavity is by getting some extra energy, this extra energy comes from the mechanical system. On the other hand, in the blue-detuned case, the pumped field is the one that gives energy to the mechanical resonator.}

\begin{figure}[H]
\begin{center}
\includegraphics[width = 0.75\textwidth]{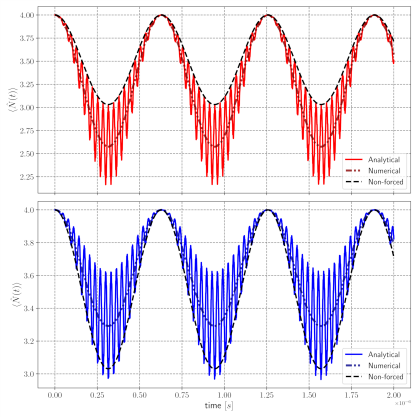}
\caption{Temporal evolution of the average phonon number with Hamiltonian parameters $\omega_c=10^9 \, \text{s}^{-1}$, $\omega_m = \omega_c/100$, $\Gamma=2$, $\alpha$ = 2, $G_0/\omega_m=0.033$, $\Omega=(\pi/20) \omega_c$. Upper panel $\omega_p=0.8 \omega_c$ (red detuned). Lower panel $\omega_p=1.2 \omega_c$ (blue detuned)}
\label{figura3-1b}
\end{center}
\end{figure}

\begin{figure}[H]
\centering
\includegraphics[width = 0.75\textwidth]{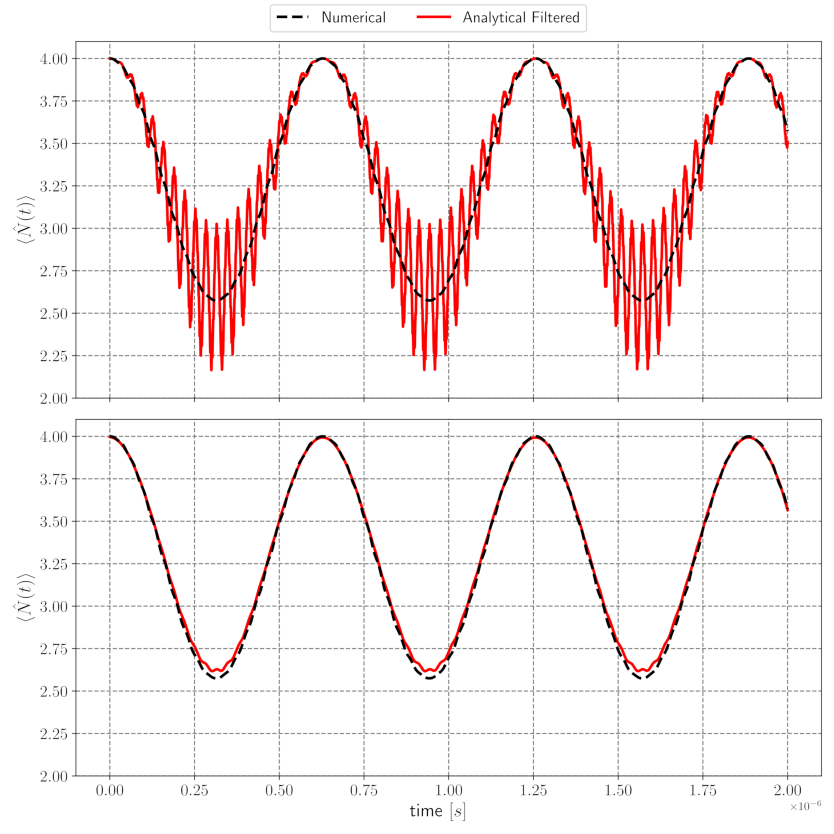}
\caption{Temporal evolution of the average phonon number with Hamiltonian parameters $\omega_c=10^9 \, \text{s}^{-1}$, $\omega_m = \omega_c/100$, $\Gamma=2$, $\alpha$ = 2, $G_0/\omega_m=0.033$, $\Omega=(\pi/20) \omega_c$, $\omega_p=0.8 \omega_c$ (red detuned) upper panel without filtering, lower panel filtered results.}
\label{filtro}
\end{figure}

\begin{figure}[H]
\begin{center}
\includegraphics[width = 0.9\textwidth]{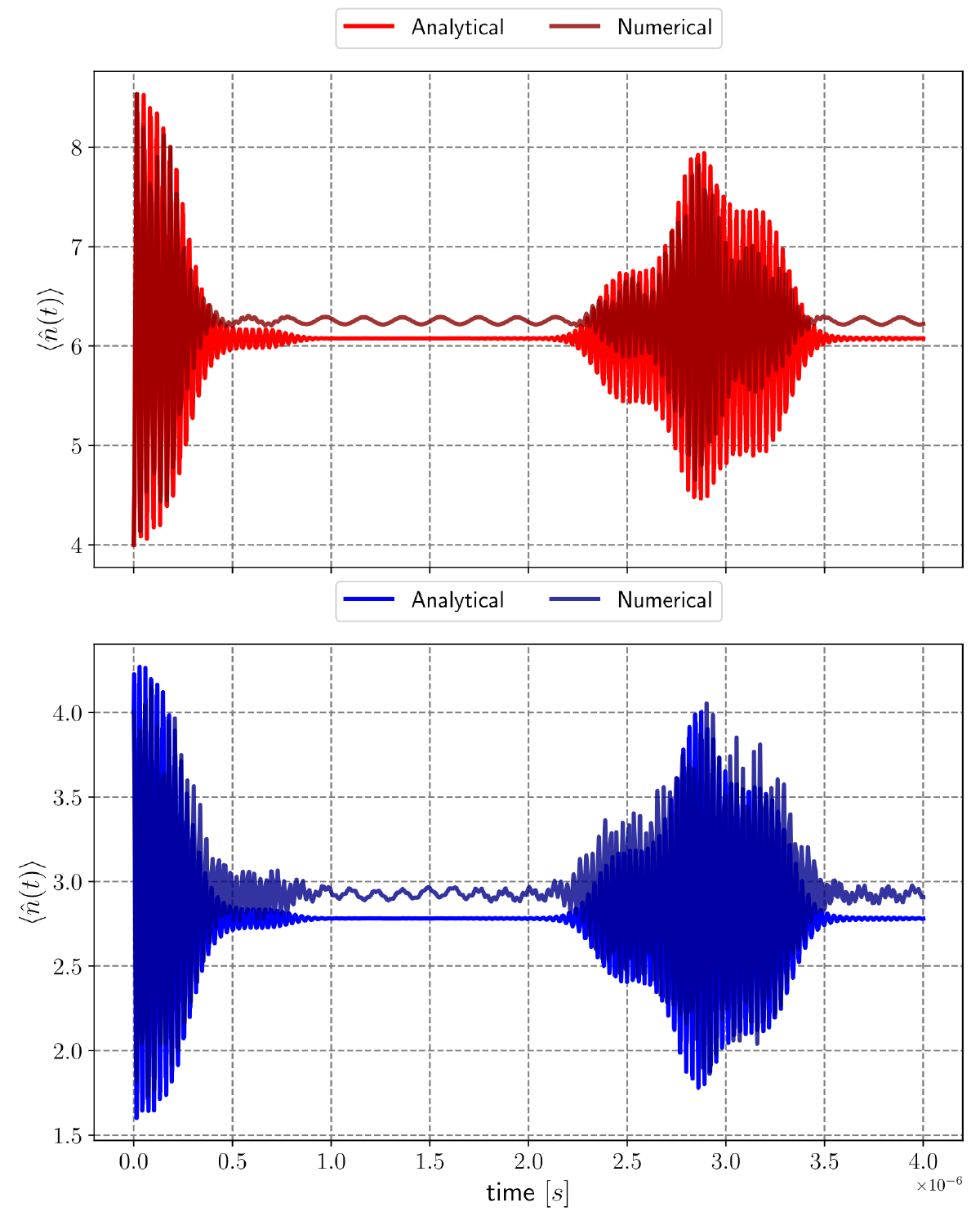}
\caption{Temporal evolution of the average photon number with Hamiltonian parameters at the strong coupling $\omega_c=10^9$, $\omega_m = \omega_c/100$, $\Gamma=2$, $\alpha$ = 2, $G_0/\omega_m=0.33$, $\Omega=(\pi/20)\omega_c$. Upper panel $\omega_p=0.8 \omega_c$ (red detuned, dark color is the numerical solution). Lower panel $\omega_p=1.2 \omega_c$ (blue detuned, dark color is the numerical solution)}
\label{figura3a}
\end{center}
\end{figure}

\begin{figure}[H]
\begin{center}
\includegraphics[width = 0.75\textwidth]{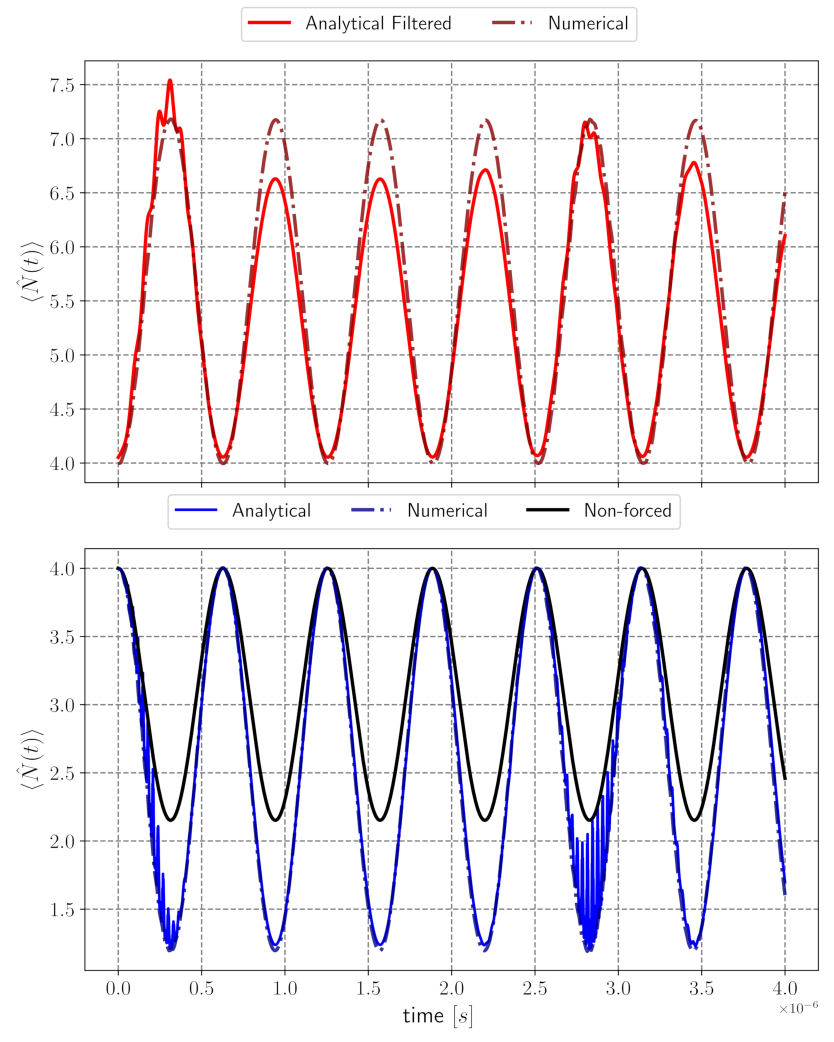}
\caption{Temporal evolution of the average phonon number with Hamiltonian parameters at the strong coupling regime $\omega_c=10^9 \, \text{s}^{-1}$, $\omega_m =\omega_c/100$, $\Gamma=2$, $\alpha$ = 2, $G_0/\omega_{m}=0.33$, $\Omega=(\pi/20) \omega_c$. Upper panel $\omega_p=0.8 \omega_c$ (red detuned). Redline filtered analytic results, brown broken-line numerical results. Lower panel $\omega_p=1.2 \omega_c$ (blue detuned). Blue line analytic results with no filtering, blue broken-line numerical results. In black are the numerical results for the non-forced system.}
\label{figura3b}
\end{center}
\end{figure}

In Fig.~\ref{figura3-1b}, we show the temporal evolution for the average phonon number; in the upper panel the forcing frequency is $\omega_p=0.8\omega_c$ (red detuned), and in the lower panel, it is $\omega_p=1.2\omega_c$ (blue detuned). The initial value for the average phonon number is $\bar N=\langle \hat N\rangle= |\Gamma|^2=4$. In these figures, we also show the exact results obtained for an optomechanical system with no forcing term (black dashed line), the numerical results for the forced optomechanical system (dotted darker blue or red line), and the results obtained with the analytic method (solid blue or red line); the Hamiltonian parameters are the same as those used for Fig.~\ref{figura3-1a}. In both cases (blue detuned and red detuned), the average value of the phonon number is a decreasing function of time, with an envelope oscillating with the period of the mechanical oscillator $T_m=2\pi/\omega_m$. The analytic results also show rapid oscillations, with period $T_{\Delta}$ due to the average photon number evolution (see Eq.~\eqref{eq:heis}). The main difference between the red and blue detuned cases is that the effect of the forcing term in the red detuned case is a larger decrease of the average phonon number concerning the nonforced system, while in the blue detuned case, the average phonon number for the forced system decreases less than the decrease for the nonforced system. In the red-detuned case, we can see that the average photon number is an increasing oscillating function, and the average phonon number is a decreasing oscillating function that takes values smaller than its values for the nonforced system. For the blue detuned case, the average photon number is a decreasing oscillating function, and the average phonon number is a decreasing oscillating function that takes values above those of the nonforced system. For this set of Hamiltonian parameters, the red-detuned and the blue-detuned cases show a cooling behavior of the mechanical oscillator.  
In Fig.~\ref{filtro}, we show the results obtained when we filter the rapid oscillations obtained in the analytic approximation for the phonon number operator and compare them with the numerical calculation.  Recall that in our analytic approach, we have replaced the operators appearing in the exponents in the expression for the interaction Hamiltonian (see Eq.~\eqref{eq:Hinteraction}) by an average taken between coherent states; then, as a consequence of this approximation, 
interference terms may not be present. However, we demonstrate that when we take an average of these spurious oscillations, the analytic results converge to the numerical results, see the lower panel of Fig.~\ref{filtro}.
In Fig.~\ref{figura3a} upper panel, we plot the temporal evolution of the average photon number obtained with the analytical method described above for $\omega_p=0.8\omega_c$ (red lines), and the results obtained using a purely numerical calculation with the full Hamiltonian (dark red lines) using Qutip \cite{Johansson2012}, for a strong coupling constant $G_0/\omega_{m}=0.33$, ten times larger than that used in the previous figures; as before, the initial value is $ \bar n(t_0) = 4$. At the beginning of the evolution $\bar n(t)$ increases up to $\bar n=9$, with rapid oscillations going down to $\bar n=4$; after a time of the order of $5\times 10^{-7}$ seconds, we notice the appearance of a first collapse, similar to those seen in the Jaynes-Cummings model, the oscillations cease with $\bar n(t)$ taking a  constant value around $\bar n=6$. The length of the collapse is $\tau\simeq 2\times 10^{-6}$ seconds and after that, we see a new revival in the oscillations, its duration being of the same order of magnitude as that of the collapse. In the lower panel, the forcing frequency $\omega_p=1.2\omega_c$ (blue detuned); the analytic results are shown in blue, and the numerical results in dark blue. Collapses and revivals are also present in this case, the main difference between both cases is that with red detuning $\bar{n}(t)$ oscillates above its initial value, while for blue detuning it oscillates below its initial value, conduct we had already seen for the case of a small $G_0/\omega_m$. The duration of the collapses and revivals is similar in both cases. We can see a good qualitative agreement between the numerical and the analytic calculations.
In Fig.~\ref{figura3b}, we see the temporal evolution of the average phonon number using the same Hamiltonian parameters as those of Fig.~\ref{figura3a}. The upper panel corresponds to red detuning and the lower panel to blue detuning; in both cases, the initial average value of the phonon number operator is 4. In the red detuning case, we show the results obtained with the analytic method after averaging the fast oscillations due to the revival of the average photon number in the red line, and in the brown broken line the numerical results obtained with the total Hamiltonian. We notice that the filtered results have a reasonable agreement with the numerical results, the main difference is seen in the amplitude of the oscillation in the temporal region of the photon collapse; the frequency of the oscillations is the same all along the temporal evolution shown in the figure. We see that the average phonon number has an increasing oscillation in the range $4\leq \langle \hat N(t)\rangle \leq 8$ so that the average number of phonons increases from its initial value.  In the lower panel, we show in blue full-line the analytic results with no filtering applied, the numerical results in blue line-dot, and black line the results for the non-forced system. The non-forced system shows a decrease in the average phonon number in the range $2< \langle \hat N(t)\rangle\leq 4$, with the forcing present, the average phonon number attains even smaller numbers in the range $1<\langle \hat N(t)\rangle\leq 4$, so that, with blue detuning,  the forcing term cools the mechanical oscillator. The agreement between the analytic and the numerical results is relatively good in the region of the collapses of the photon number, the main differences are present in the areas of the revivals of the photon number.
The presence of collapses and revivals in an optomechanical system, similar to those shown in the Jaynes-Cummings model of quantum optics, was predicted in \cite{Dobrindt2008}. The analogy with quantum optics can be understood when the optomechanical coupling constant relies on the strong coupling regime ($G_0>\kappa, \gamma_m$), where $\kappa$ is the amplitude for the cavity decay rate, and $\gamma_m$ is the rate at which the oscillator exchanges phonons with the environment since to have coherent dynamics it is necessary that the time scale, in which the optomechanical interaction takes place, be smaller than the decoherence time scales in the system.

\subsection{Mandel parameter and linear entropy}
Consider now the Mandel parameter for the field defined as~\cite{Mandel_79}
\begin{equation}\label{eq:mandel}
Q = \frac{\langle \hat n^2(t)\rangle -\langle \hat n(t) \rangle^2}{\langle \hat n(t) \rangle}.
\end{equation}
When the averages are taken between coherent states $Q=1$, the distribution is Poissonian. \textcolor{black}{In the case of the Fock states $Q=0$, and the photon distribution is sub-Poissonian. Note that $Q<1$ is a sufficient condition for the cavity to be in a nonclassical state of light, but for $Q>1$ no direct conclusion can be drawn about its nonclassicality~\cite{agarwal2012quantum}.}

In Eq.~\eqref{eq:photonsHeis} we presented the photon number operator in the Heisenberg representation $\hat n(t)$, so that
\begin{equation}\label{eq:ncuadrada}
\hat n^2(t)= \hat U^{\dagger}\hat n\hat U \hat U^{\dagger}\hat n \hat U =(\hat n-\beta_2\hat a+\beta_1\hat a^{\dagger}-\beta_1\beta_2)(\hat n-\beta_2\hat a+\beta_1\hat a^{\dagger}-\beta_1\beta_2),
\end{equation}
with $\hat n$ the photon number operator in the Schrödinger picture. When we take the average values given in Eq.~\eqref{eq:mandel} between initial coherent states and make use of Eqs. \eqref{eq:photonsHeis} and \eqref{eq:ncuadrada}, we obtain $Q=1$, a value that is preserved along the evolution. \textcolor{black}{This result resembles the one that is obtained during the generation of Schr\"odinger cat states using a nonlinear medium.  Under the influence of a nonlinear Kerr-like Hamiltonian, a cavity field evolves into a highly non-classical state of light, the cat state $|\psi_{\rm ys}\rangle\equiv(|\alpha\rangle+i|-\alpha\rangle)/\sqrt{2}$, also known as the Yurke-Stoler state~\cite{yurke86}. This cat state is an eigenstate of the square of the annihilation operator, $\hat a^2|\psi_{\rm ys}\rangle=\alpha^2 |\psi_{\rm ys}\rangle$, and satisfies $Q=1$~\cite{Gerry2004}. }

\begin{figure}[H]
\begin{center}
\includegraphics[width = 0.9\textwidth]{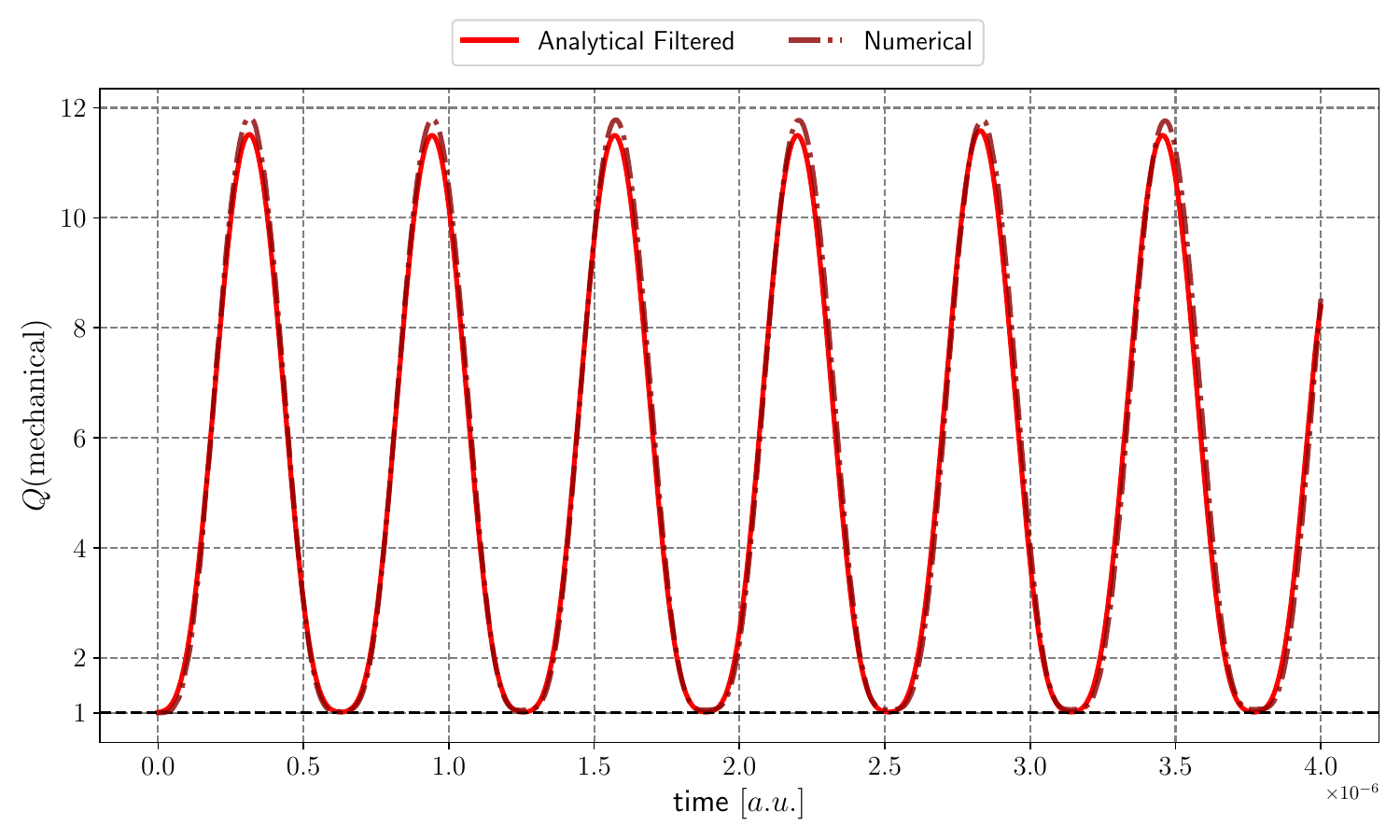}
\caption{Temporal evolution of the Mandel parameter for the mechanical oscillator with Hamiltonian parameters at the strong coupling $\omega_c=10^{9}\,\text{s}^{-1}$, $\omega_m =\omega_c/100$, $\omega_p=0.8\omega_c$, $\Gamma=2$, $\alpha$ = 2, $G_0/\omega_{m} = 0.33$, $\Omega=(\pi/20) \omega_c$}
\label{figura-Mandel}
\end{center}
\end{figure}

We can also evaluate the Mandel parameter for the mechanical oscillator
\begin{equation}
Q_{M}= \frac{\langle \hat N^2(t)\rangle -\langle \hat N(t) \rangle^2}{\langle \hat N(t) \rangle},
\end{equation}
where the average number of phonons is given by Eq.~\eqref{eq:heis}, and
\begin{equation}
\langle \hat N^2(t) \rangle = \langle \Psi(t)|\hat N^2 |\Psi(t)\rangle = \langle \Psi(t_0)|\hat N^2(t)|\Psi(t_0)\rangle .
\end{equation}
In the Heisenberg representation the operator $\hat N^2(t)$ is given by
\begin{equation}
\hat N^2(t) = \left(\hat N +(\alpha_3 \hat b^{\dagger}+\alpha_3^{*}\hat b)\hat n(t) +|\alpha_3|^2\hat n^2(t)\right)^2,
\end{equation}
and its average value between coherent states $|\Gamma\rangle$ is
\begin{align}
\langle\hat{N}^2(t)\rangle=&\langle\hat{N^2}\rangle+4\Re{\alpha_3\Gamma^*\left(|\Gamma|^2+\frac{1}{2}\right)}\langle\hat{n}(t)\rangle
\nonumber \\ &
+2\left(\Re{(\alpha_3\Gamma^*)^2}+|\alpha_3|^2\left(2|\Gamma|^2+\frac{1}{2}\right)\right)\langle\hat{n}^2(t)\rangle
\nonumber \\ &
+4|\alpha_3|^2\Re{\alpha_3\Gamma^*}\langle\hat{n}^3(t)\rangle+|\alpha_3|^4\langle\hat{n}^4(t)\rangle.
\end{align}
With these expressions, we can compute the Mandel parameter for the mechanical oscillator. The result is found in Fig. \ref{figura-Mandel}, after filtering the rapid oscillations (red line), along with the result of the purely numerical computation (brown broken line); we see a very good agreement between them. We also see that the Mandel parameter for the mechanical oscillator is larger  than one, corresponding to a super-Poissonian statistic.\\
Consider now the linear entropy for the subsystem $x$, where $x$ is the label for the mirror or the cavity, which is given by $S^{(x)}= 1-\mathrm{Tr_{x}}[\rho_x^2]$. The reduced density matrix for the mechanical oscillator is
\begin{equation}
\rho_m(t) = \mathrm{Tr_c}[\rho] = e^{-|\alpha+\beta_1|^2}\sum_p \frac{|\alpha+\beta_1|^{2p}}{p!}|\Gamma_p(t)\rangle \langle \Gamma_p(t)|.
\end{equation}
From this expression, we obtain
\begin{equation}
\mathrm{Tr_m}[\rho^2_m(t)]= e^{-2|\alpha+\beta_1|^2}\sum_{p,q}\frac{|\alpha+\beta_1|^{2p}}{p!}\frac{|\alpha+\beta_1|^{2q}}{q!}e^{-|\Gamma_p(t)-\Gamma_q(t)|^2},
\end{equation}
and we can evaluate the linear entropy for the mechanical oscillator.

\begin{figure}[H]
\begin{center}
\includegraphics[width = 0.9\textwidth]{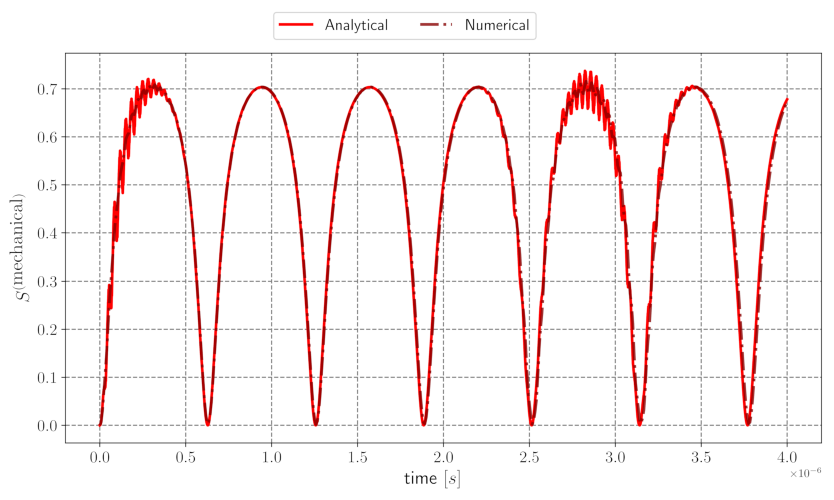}
\caption{Temporal evolution of the linear entropy for the mechanical oscillator with Hamiltonian parameters at the strong coupling $\omega_c=10^{9} \, \text{s}^{-1}$, $\omega_m = \omega_c/100$, $\omega_p=0.8\omega_c$, $\Gamma=2$, $\alpha$ = 2, $G_0/\omega_{m}=0.33$, $\Omega=(\pi/20) \omega_c.$}
\label{figura12}
\end{center}
\end{figure}

We see in Fig. \ref{figura12} the evolution of the linear entropy for the mechanical oscillator. It presents an oscillatory conduct with the period of the mechanical oscillator, with small oscillations with the frequency of the detuning superimposed. These are evident in the region of the revivals, while in the region of the collapses, the linear entropy oscillates only with frequency $\omega_m$.  At times $t_n=2\pi n/\omega_m$, the entanglement between the field and the mechanical oscillator is zero, the system returns to a pure state and the linear entropy goes to zero. \textcolor{black}{At times $t_n=(2n + 1)\pi/\omega_m$, the entanglement is maximum and the linear entropy attains its maximum value, which is not 1.} In this figure, we have not filtered the rapid oscillations, these can be seen in the regions of the revivals of the photon number operator.

\section{Wigner function}\label{wigner_section}
Having determined the expectation values of several observables, and the degree of mixedness and entanglement on the intracavity field and the mechanical mirror, via the linear entropy, we now show a representation of the wavepackets evolution in phase space using the Wigner function. We choose to make a comparison between a purely numerical solution of the Hamiltonian \eqref{Hforced}, and the solution using the wavefunction \eqref{wavefunc}, derived from the approximated interaction Hamiltonian \eqref{HApprox}. To begin, we recall the definition of the continuous
\begin{equation}
W_{\hat{\rho}}(q, p) = \frac{1}{\pi}\int\limits_{-\infty}^{\infty}dx\,\exp\left(-2\mathrm{i}xp\right)\bra{q-x}\hat{\rho}\ket{q+x},
\end{equation}
 and discrete Wigner function
\begin{equation}
\tilde{W}_{\hat{\rho}}(q, p) = \frac{1}{N}\sum\limits_{n}\exp\left(-\frac{4\mathrm{i}\pi np}{N}\right)\bra{q-n}\hat{\rho}\ket{q+n},
\end{equation}
 of a density matrix, $\hat{\rho}$ \cite{Schleich2001, Bianucci2002}, where $q$ and $p$ are the position and canonical momentum in the phase-space, and $N$ is the dimension of the discrete and finite Hilbert space used to spawn the phase-space coordinates; either way, the procedure to obtain the Wigner functions for the comparison is similar. In the first case, a complete numerical solution of the initial forced optomechanical Hamiltonian \eqref{Hforced} is given to an ordinary differential equation solver in Python's QuTip \cite{Johansson2012} and Julia's QuantumOptics.jl \cite{Kramer2018}, returning the evolved density matrix; i.e., there is no approximation in the dynamical equations, just numerical considerations for the discretization of the Fock space. In the second case, we treat the approximated Hamiltonian \eqref{HApprox}, first numerically solving their time-dependent coefficient equations $\dot{\beta}_{i}$ in \eqref{eqs0350}, \eqref{eqs0360}, \eqref{eqs0370}, and putting them in the wavefunction \eqref{wavefunc}. For both cases, we use the built-in Wigner function routine in the numerical packages to obtain the respective phase-space representation.

\begin{figure}[H]
\centering
\includegraphics[width = 0.8\textwidth]{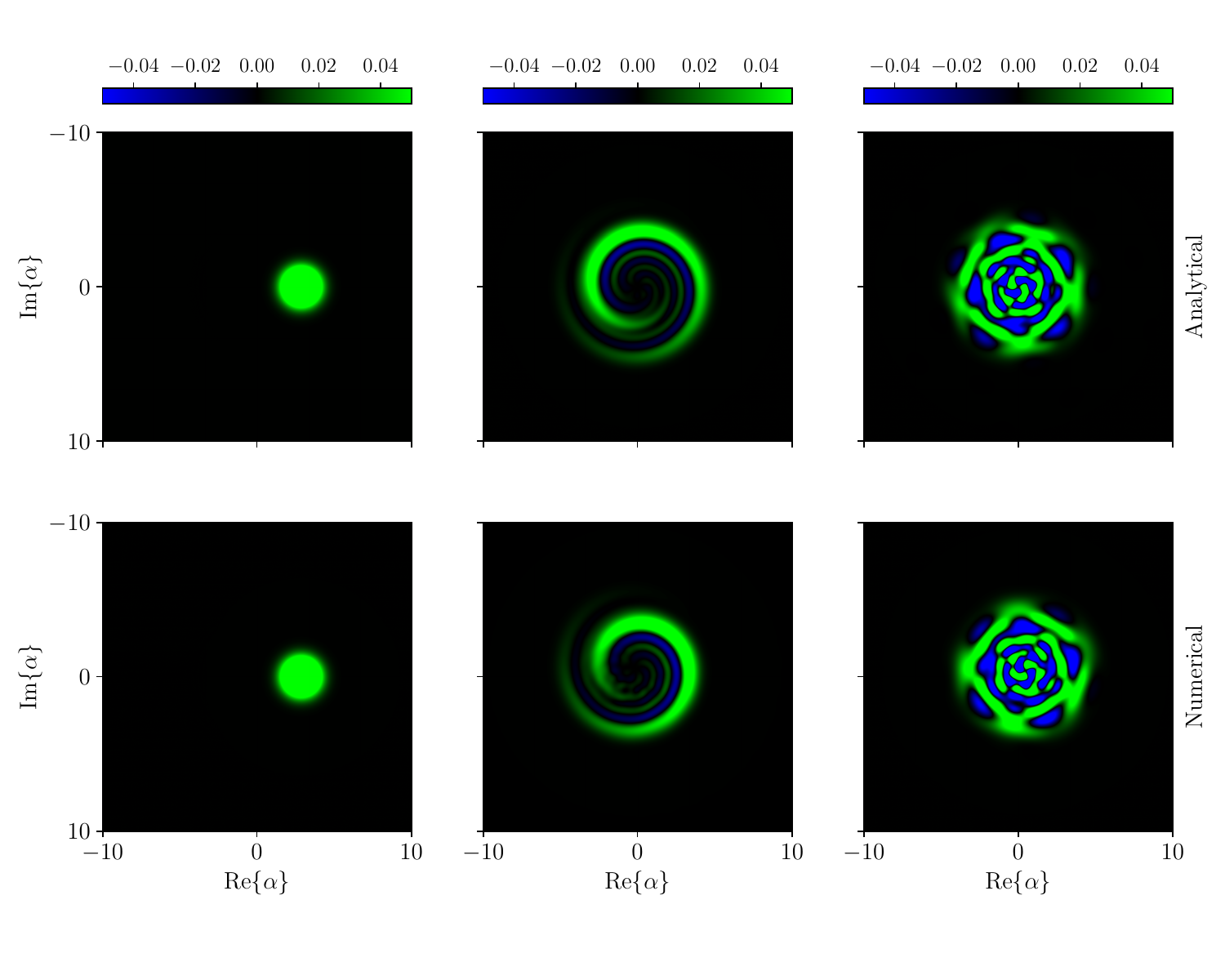}
\caption{Plot of the Wigner function for the cavity field evolution. The initial eigenfunction is a coherent state $\ket{\alpha = 2}$. The selected snapshots are where the entanglement between the field and the mechanical mirror is a minimum or a maximum; that is, from left to right, $t_{n} = \left\{0, \frac{\pi}{\omega_{m}}, \frac{2\pi}{\omega_{m}}\right\}$. In the row above, we have the analytical solution given by the wavefunction \eqref{wavefunc}, and in the row below, we have the numerical solution solving the Schrödinger equation for the original Hamiltonian \eqref{Hforced}}. \label{wigner_field}
\end{figure}

\begin{figure}[H]
\centering
\includegraphics[width = 0.85\textwidth]{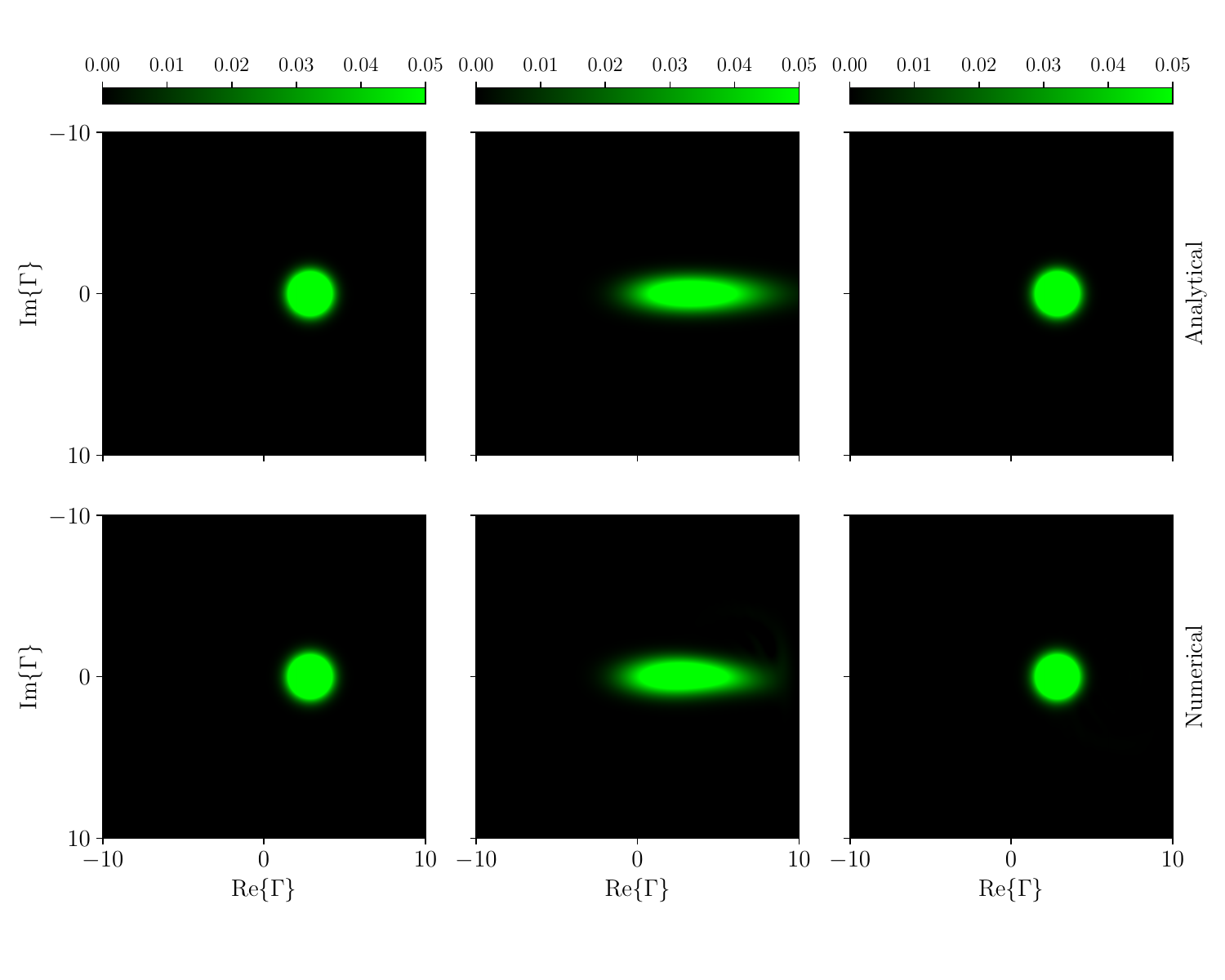}
\caption{Plot of the Wigner function for the mechanical mirror evolution. The initial eigenfunction is a coherent state $\ket{\Gamma = 2}$. The selected snapshots are at times where the entanglement between the field and the mechanical mirror is a minimum or a maximum; that is, from left to right, $t_{n} = \left\{0, \frac{\pi}{\omega_{m}}, \frac{2\pi}{\omega_{m}}\right\}$. In the row above, we have the analytical solution given by the wavefunction \eqref{wavefunc}, and in the row below, we have the numerical solution solving the Schrödinger equation for the original Hamiltonian \eqref{Hforced}.}
\label{wigner_osc}
\end{figure}

In Figs.~\ref{wigner_field} and \ref{wigner_osc}, we show the Wigner function for the field and the mechanical oscillator, respectively. The parameters are carried on from the set used above for the linear entropy, i.e., $\omega_{c} = 10^{9}\,\textrm{s}^{-1}$, $\omega_{m} = \omega_{c}/100$, $\omega_{p} = 0.8\omega_{c}$, $G_{0}/\omega_{m} = 0.33$, $\Omega = (\pi/20)\omega_{c}$, and the initial coherent wavefunctions with $\alpha = 2$, and $\Gamma = 2$. In each figure, the row above shows the Wigner function using the analytical solution given by the wavefunction~(\ref{wavefunc}), and the row below shows the Wigner function using the numerical solution of the Schrödinger equation using the original Hamiltonian~(\ref{Hforced}). From left to right, the snapshots correspond to the times $t_{n} = \left\{0, \frac{\pi}{\omega_{m}}, \frac{2\pi}{\omega_{m}}\right\}$ where minimum, maximum and minimum entanglement is present. From Fig.~\ref{wigner_field}, the visual verification reveals that the field evolves in a non-classical form, as a quantum harmonic oscillator whose dynamics have some degree of complexity, giving place to phase space interferences and negative values in the Wigner function; we can see a very good agreement between both calculations, an indication of the accuracy of the approximate methodology we used. \textcolor{black}{In Fig.~\ref{wigner_osc}, we present the Wigner function for the mechanical oscillator, which is positive at all times and shows a squeezed profile in the momentum direction, i.e., when the entanglement is maximum at $t=\pi/\omega_m$.} It can be seen that there are very small differences between the purely numerical calculations and the analytic ones due to the approximations made in the analytic treatment.

\textcolor{black}{
Finally, we emphasize that we are not treating each sub-system separately; in fact, our work aims to obtain the time evolution operator of the entire driven optomechanical system and analyze the corresponding evolution. For instance, Eq.~(\ref{total_U}) is the time evolution operator of the total coupled system, and Eq.~(\ref{wavefunc}) is the state vector of the total coupled system. The computation of the average photon number and average phonon number is because $\hat n$ and $\hat N$ are observables and can be experimentally obtained from local operations on the coupled system. The same idea lies behind the Mandel $Q$ parameter. In principle, one can also experimentally measure each Wigner function using the quantum-state tomography approach, which once again relies on local measurements of the field's and mechanical oscillator's quadratures, see Refs.~\cite{Wigner_PRL,RMP_Lvovsky_2009}.}

\section{Conclusions} \label{sect-conclusions}
In this work, we have obtained the exact time-evolution operator for the undriven optomechanical system, and we have evaluated the evolution of the average phonon number as a function of the effective coupling parameter $(G_0/\omega_m) |\alpha|^2$. We have seen that for small values of the effective coupling parameter the mechanical oscillator cools down, and the amount of cooling increases as we take larger values of the coupling until a point where a further increase of the coupling reverses this conduct, the cooling diminishes and we can attain a region where heating is reached. We developed an algebraic method to obtain an approximate time-evolution operator for the driven optomechanical system. The forcing term in the total Hamiltonian is responsible for the time-dependent Lie algebra in the interaction picture Hamiltonian, therefore it is necessary to make approximations to apply the standard algebraic method. The transformed creation-annihilation operators have an effective frequency, that contains operators corresponding to the mechanical oscillator and the field. Due to the presence of non-commuting operators, we cannot apply the Wei-Norman theorem \cite{Wei1964}, so we approximate the exponential by its average value taken between initial coherent states for the field and for the mechanical oscillator, obtaining an approximate interaction Hamiltonian whose time evolution operator is exact.

With this evolution operator, we evaluated the evolution of the average values of the number of photons and phonons, and we found a very good agreement between the analytic results and the numerical results for the case of the photons; for the phonons and coupling constant in the strong regime, we obtained fast oscillations that are not seen in the numerical calculation. When we take an average of these fast oscillations, the analytic and the numerical results show a good agreement. 

\textcolor{black}{To find out the statistical properties of the mechanical oscillator states, we evaluated its Mandel $Q$ parameter, and we found that they are super-Poissonian. We also computed the linear entropy, and we found that periodically, the entropy goes to zero, indicating that at such times, the system attains a pure state; we also found that at other periodical times, the entropy achieves its maximum and the system is entangled. Finally, we calculated the Wigner function for the field and the mechanical oscillator using the approximate evolved state and also using a purely numerical calculation. For the field, we found a clear non-classical behavior, since the Wigner function attains negative values; for the mechanical oscillator, the Wigner function is positive at all times but shows squeezing at the times of largest entanglement with the field corresponding to non-classical behavior.}

It is worth mentioning the qualitative agreement of the Wigner functions obtained from the methodology developed in this work, and the purely numerical one.\\ 

\noindent {\bf Acknowledgements:} 
J.R. and L.M.-D. acknowledge partial support from DGAPA-UNAM project IN109822. 
\textcolor{black}{R.R.-A. thanks DGAPA-UNAM, M\'exico for support under Project No. IA104624.}
A.R. U. acknowledges postdoctoral support from CONAHCyT and DGAPA-UNAM. 

\bibliographystyle{unsrt}
\bibliography{bib}

\end{document}